\documentclass[%
 reprint,
 amsmath,amssymb,
 aps,
]{revtex4-1}

\usepackage{graphicx}
\usepackage{dcolumn}
\usepackage{bm}
\usepackage{braket}
\usepackage{amsmath}
\usepackage{amssymb}
\usepackage{feynmp}
\usepackage{graphics}
\usepackage{color}
\begin{document}


\title{Metal-insulator transition in transition metal dichalcogenide heterobilayer moir\'e superlattices}

\author{Nicol\'as Morales-Dur\'an}
\affiliation{Department of Physics, University of Texas at Austin, Austin, Texas, 78712, USA}
\author{Pawel Potasz}%
\affiliation{Department of Physics, University of Texas at Austin, Austin, Texas, 78712, USA}
\affiliation{Department of Physics, Wroclaw University of Science and Technology, 50-370 Wroclaw, Poland} 
\author{Allan H. MacDonald}%
\affiliation{Department of Physics, University of Texas at Austin, Austin, Texas, 78712, USA}%




\date{\today}
\begin{abstract}
Moir\'e superlattices formed in two-dimensional semiconductor heterobilayers provide a new realization of Hubbard model physics in which the number of electrons per effective atom can be tuned at will.  We report on an exact diagonalization study of the electronic properties of half-filled narrow moir\'e bands in which correlation strengths are varied by changing twist angles or interaction strengths.  We construct a phase diagram for the bilayer, identifying where the metal-insulator phase transition occurs, estimating the sizes of the charge gaps in the insulating phase, and commenting on the nature of the transition and the importance of sub-dominant interaction parameters.
\end{abstract}

\pacs{Valid PACS appear here}
\maketitle


\section{\label{sec:level1}Introduction}
A moir\'e superlattice is formed when two or more van der Waals layers are stacked with small differences in 
lattice constant or orientation.  When the isolated layers are semiconductors or semimetals, the electronic properties of the bilayer are accurately described by continuum models that have the periodicity of the moir\'e superlattice, thereby realizing moir\'e materials
- artificial two-dimensional crystals in which the lattice constant is on the moir\'e pattern scale.  
One of the most attractive aspects of moir\'e materials is that the longer periodicity allows the number of electrons per 
effective atom to be tuned through large ranges with electrical gates. 
When the moir\'e minibands are flat,
electronic correlations are strong and can lead to new physics.
In magic angle twisted bilayer graphene, 
for example, strong correlations are manifested by insulating states 
surrounded by superconducting domes \cite{Jarillo,Jarillo2,Efetov, Dean1}, and quantum anomalous Hall ferromagnets \cite{GoldhaberGordon,Young}.

\begin{figure}[h!]
\includegraphics[width=\linewidth]{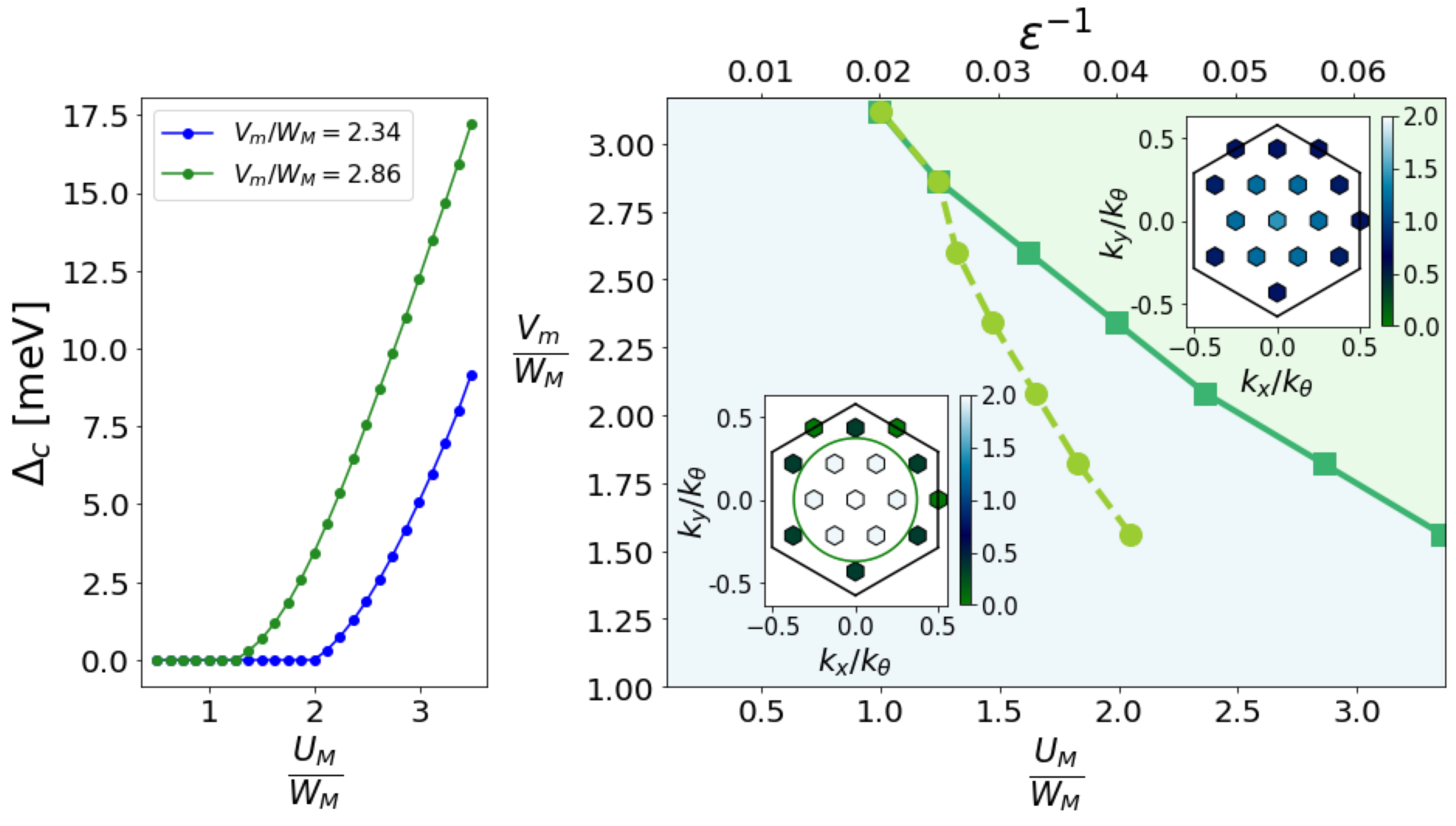}
\caption{(a) Extrapolated charge gap {\it vs} $U_M/W_M$,
for two values of the normalized moir\'e potential depth $V_m/W_M$. Here $W_M=\hbar^2/(m^*a_M^2)$ is the moir\'e kinetic-energy 
scale and $U_M=e^2/(\epsilon a_M)$ is the moir\'e interaction energy scale. 
(b) Phase diagram for a TMD heterobilayer {\it vs.} $V_m/W_M$ and $U_M/W_M$, the metallic and insulating phases are indicated in blue and green respectively. Lines connect points estimated to be on the metal-insulator phase transition boundary calculated from exact diagonalization (dark green) and Hartree-Fock (light green).
The insets show the typical momentum distribution functions of each phase on a finite-size discrete momentum space mesh. 
The Fermi surface of the metallic phase is schematically indicated by a circle. 
The top axis indicates the values of dielectric constant, $\epsilon$, 
corresponding to a given $U_M/V_M$ for a heterobilayer twisted by $\theta=2.5^{\circ}$.}
\label{fig:Chargegap}
\end{figure}

In this article we report on an exact diagonalization study of the moir\'e superlattices 
formed in transition metal dichalcogenide (TMD) heterobilayers
in which correlated insulators and Wigner crystal
states have already been observed \cite{HubbardCornell,Columbia,Berkeley, ETH, Cornell2,Cornell3,TMDcorrelated1}. 
In heterobilayer systems, which have different two-dimensional semiconductors
on opposite sides of the junction, there is an interval of energy near the band extremum within which
carriers are localized in one of the two layers.  For example, for WSe$_2$ heterobilayers 
formed with either MoSe$_2$ or MoS$_2$, the carriers at the top of the valence band are 
localized in the WSe$_2$ layer, but experience a periodic potential due to the moir\'e pattern.
Spin-valley locking in WSe$_2$ then leads to low-energy physics described by a Hubbard-like
model in which spinful electrons experience a periodic potential whose extrema form a triangular 
lattice \cite{FengchengHubbard, LiangFuTransfer}.  
We limit our attention to the case of 
one-electron per moir\'e period and focus on
the metal to insulator phase transition (MIT) \cite{Imada,Mott}  
that is expected to occur when interactions become strong compared to moir\'e miniband widths. 

The bilayer is described by a continuum model that depends on moir\'e potential depth $V_m$ and on the moir\'e period 
$a_M$ (or equivalently the twist angle), which determine the kinetic energy scale $W_M$ and interaction energy scale $U_M$. 
Our main results are summarized by the phase diagram in Fig.~\ref{fig:Chargegap}, which is controlled by the 
dimensionless parameters $V_m/W_M$ and $U_M/W_M$. We find that the metal-insulator phase transition 
points can be readily identified by calculating the charge gap $\Delta_c$ {\it vs.} $U_M/W_M$ for fixed 
$V_M/W_M$, as shown in Fig. \ref{fig:Chargegap}(a). Repeating these calculations at different modulation strengths $V_m$
yields the phase diagram shown in Fig. \ref{fig:Chargegap}(b). In order to emphasize the importance of a non-mean-field theory treatment of the metal-insulator phase transition, we have included an estimate for transition line obtained from the Hartree-Fock method applied to a groundstate without broken translational symmetry. The insulating state is favored, as expected, at large $V_m$ and $U_M$ but its stability is overestimated by the Hartree-Fock calculation. Momentum-state occupation-number distribution functions $\langle \Psi_{GS}| n_{{\bf{k}}}|\Psi_{GS}\rangle$ 
where $|\Psi_{GS} \rangle$ is the many-body ground state, plotted as insets in Fig.~\ref{fig:Chargegap}(b),
clearly distinguish the two states by the presence or absence of a Fermi surface (schematically represented as a green circle).
These numerical results clearly indicate that a MIT occurs at half-filling in moir\'e materials,
demonstrating that they are an attractive platform for searches for superconductivity in doped Mott-insulators,
and spin-liquid states on the insulating side of metal-insulator phase transitions.  Below we first explain the 
technical details of our calculations and then discuss their implications.  

\section{\label{sec:level2} Moir\'e band model}
The moir\'e Hamiltonian of twisted TMD heterobilayer valence bands is \cite{FengchengHubbard}
\begin{equation}
\label{ContinuumHamiltonian}
    \mathcal{H}=-\frac{\hbar^2}{2 m^*}{\bf{k}}^2+\Delta({\bf{r}}),
\end{equation}
where $\Delta({\bf{r}})$ is an external potential with moir\'e periodicity. Experimental \cite{Effectivemass2} and theoretical \cite{Effectivemass3,Effectivemass4} values for the effective mass of valence band monolayer WSe$_2$ vary; here we take  $m^*=0.35\,m_0$. 
In the dominant harmonic approximation $\Delta({\bf{r}})=2V_m\sum_{j=1}^{3}\cos({\bf{b}}_j\cdot{\bf{r}}+\psi)$,
where ${\bf{b}}_j=k_{\theta}(\cos(2\pi j/3),\sin(2\pi j/3))$ and 
$k_{\theta}=4\pi/(\sqrt{3}a_M)$.
In this approximation the moir\'e modulation potential  
is completely characterized by strength ($V_m$) and shape ($\psi$) parameters. 
The potential strength $V_m$ depends on heterobilayer and, when strain effects are accounted for, also on twist angle. 
The shape parameter $\psi$ controls the relative depth of potential extrema locations and,
as shown in Ref. \cite{LiangFuTransfer}, strongly influences the strength of particle-hole asymmetry relative
to the half-filled moir\'e band case considered in this work.
For concreteness we choose the value $\psi=-94^\circ$, estimated from \textit{ab initio}
calculations for WSe$_2$/MoSe$_2$ in Ref. \cite{FengchengHubbard}.
For this $\psi$ the valence band potential has a single maximum centered 
at the AA positions of the moir\'e superlattice.

\begin{figure}[h!]
\includegraphics[width=\linewidth]{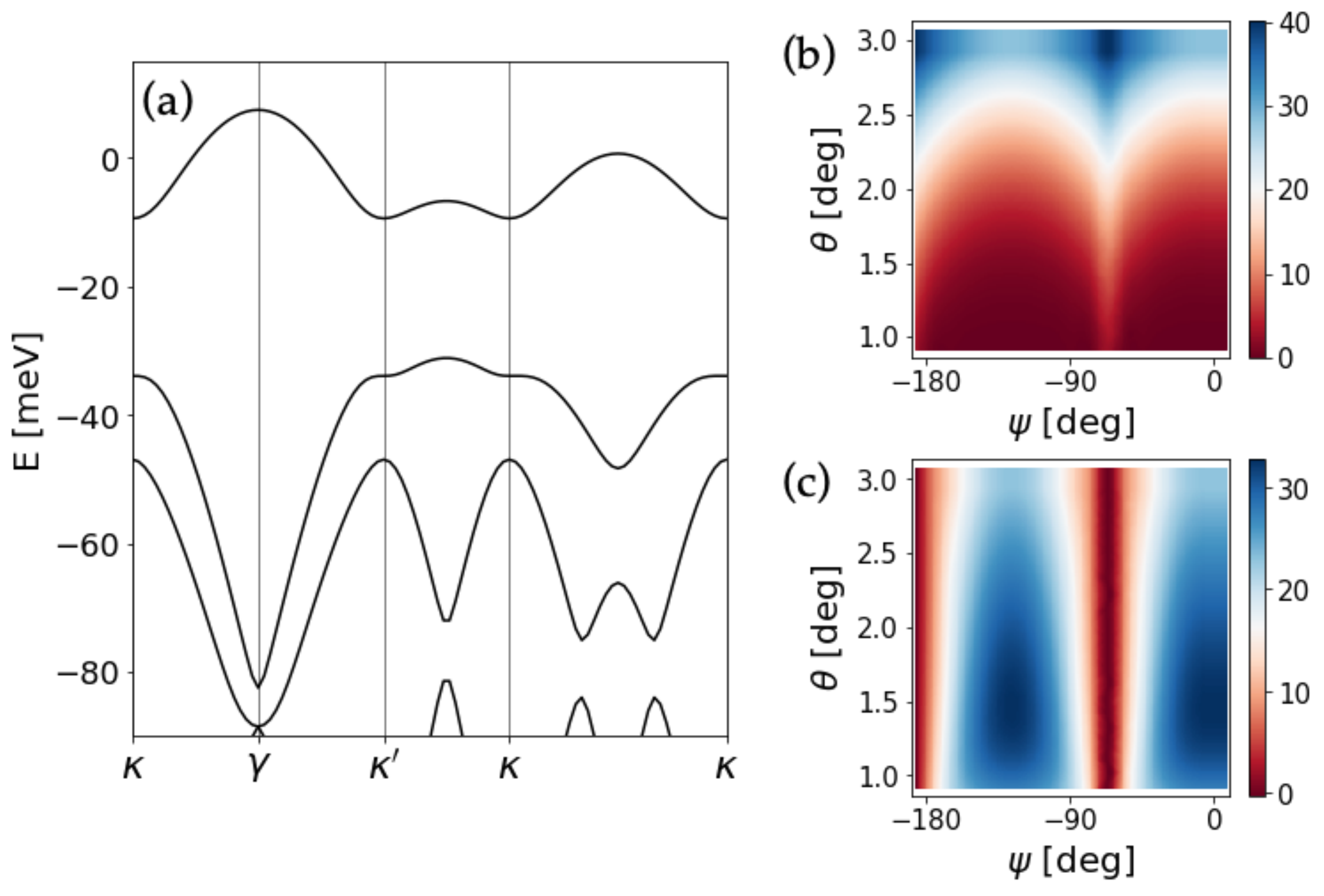}
\caption{(a) WSe$_2$/MoSe$_2$ moir\'e band structure for twist angle $\theta=2.5^{\circ}$ and shape $\psi=-94^{\circ}$, showing the topmost isolated band. 
(b) The bandwidth of the topmost band and (c) the energy gap to the first remote band {\it vs. } $\theta$ and $\psi$. In (c), the blue regions correspond to triangular lattice 
generalized Hubbard models and the red regions to honeycomb lattice generalized Hubbard models. 
Modulation strength $V_m=11$ meV was used in all illustrated calculations.}
\label{fig:BandStructure}
\end{figure}

An example of the moir\'e minibands obtained by diagonalizing the Hamiltonian in Eq. \eqref{ContinuumHamiltonian} 
in a plane wave basis is shown in Fig. \ref{fig:BandStructure}(a). 
Clearly, the topmost valence moir\'e band is well-separated and flat in this case.
The width of the topmost moir\'e band, and the energy gap to the second moir\'e band 
are plotted as a function of twist angle and shape parameter in  
Fig. \ref{fig:BandStructure}(b) and \ref{fig:BandStructure}(c) respectively. 
The width increases with twist angle but is almost $\psi$-independent except near 
$\psi=-60^\circ$ and $\psi=-180^\circ$, where the bands broaden. 
This property is explained by Fig. \ref{fig:BandStructure}(c). 
Outside of the blue regions, the topmost moir\'e miniband is not spectrally isolated, and any mapping to a one-band 
Hubbard model is inaccurate.  At both $\psi=-60^{\circ}$ and $\psi=-180^{\circ}$, the shape parameter value imposed by 
emergent symmetries in the case of $\Gamma$-valley TMD homobilayers \cite{Mattia}, the moir\'e potential has two
identical maxima that sit on honeycomb lattice sites, and it is therefore necessary to retain at least two bands to model the low-energy physics.
As $\psi$ is varied there is a smooth crossover between triangular and honeycomb lattice limits, with 
intermediate values of $\psi$ providing a realization\cite{LiangFuTransfer,LiangFu2} 
of charge-transfer insulator physics.  In this work we limit our attention to the one-band Hubbard model 
case.

Using a harmonic oscillator approximation near the highest extremum, we estimate 
that the moir\'e triangular lattice hopping parameter $t\sim \text{exp}(-\kappa\, V_m^{1/2}/\theta)$, where $\kappa$ is a constant, 
and that the on-site Coulomb interaction $U_0\sim e^2/(\epsilon a_W) \sim (e^2 \sqrt{\theta} V_m^{1/4})/\epsilon$, 
where $a_W$ is the width of the flat band Wannier function and $\epsilon$ is the static dielectric constant. 
By varying the values of $\theta$, $V_m$ and $\epsilon$, the ratios between interaction strength
and hopping can be tuned.  Experimentally, the potential depth can be varied {\em in situ}
by applying pressure \cite{Dean1, Dean2} or gate-controlled displacement fields \cite{TutucField}, 
while the dielectric environment can be modified by varying the carrier density of surrounding graphene gates \cite{BrownMATBG} 
and their separation from the active layer.
\section{\label{sec:level3} Many-Hole Hamiltonian}
Since our goal is to investigate the electronic properties of moir\'e materials, we simplify the many-body problem by projecting the continuum Hamiltonian to the Hilbert space of the topmost moir\'e miniband: 
\begin{align}
\label{ManyBodyHamiltonian}
     H&=\sum_{{\bf k}, \sigma} \epsilon_{{\bf k}, \sigma}~ c^{\dagger}_{{\bf k},\sigma} c_{{\bf k}, \sigma} \nonumber \\
    &+\frac{1}{2}\sum_{{\bf k},{\bf l},{\bf m},{\bf n}}\sum_{\sigma \sigma^{\prime}} V_{k l m n}^{\sigma \sigma^{\prime}} c^{\dagger}_{{\bf k}, \sigma}c^{\dagger}_{{\bf l},\sigma^{\prime}}c_{{\bf n}, \sigma^{\prime}} c_{{\bf m}, \sigma},
\end{align}
where $c^{\dagger}_{{\bf k},\sigma} (c_{{\bf k},\sigma})$ creates (destroys) a hole with momentum ${\bf k}$ in
valley $\sigma$, ${\bf k},{\bf l},{\bf m},{\bf n}$ are momentum labels, $\epsilon_{k,\sigma}$ is a flat valence band single particle energy, 
and $V_{k l m n}^{\sigma \sigma^{\prime}}$ is a two-particle matrix element 
\begin{align}
    V_{k l m n}^{\sigma \sigma^{\prime}}=\langle {\bf k},\sigma;{\bf l},\sigma^{\prime}|V|{\bf m},\sigma;{\bf n},\sigma^{\prime} \rangle.
\end{align}
The Coulomb long-range interaction is given by $V=e^2/\epsilon|{\bf r}_1-{\bf r}_2|$  and $\epsilon^{-1}$ is an interaction strength 
parameter related to the two-dimensional system's three-dimensional dielectric environment. The  $V_{k l m n}^{\sigma \sigma^{\prime}}$ matrix elements are sensitive to the tails of the flat band wavefunctions 
at positions away from their maxima in the moir\'e unit cell. The size of these tails is sensitive to the confinement 
potential at lattice sites, which is weaker in the moir\'e material case than in atomic lattices. 
For small twist angles the interaction physics is expected to be accurately described by a model with only on-site interactions.
For larger angles, however, longer range Coulomb interaction and non-local terms become more important (see Supplemental Material for 
further comment \cite{Supplemental}).

\section{\label{sec:level4} Metal-insulator transition}

\begin{figure}[h!]
\centering
\includegraphics[width=\linewidth]{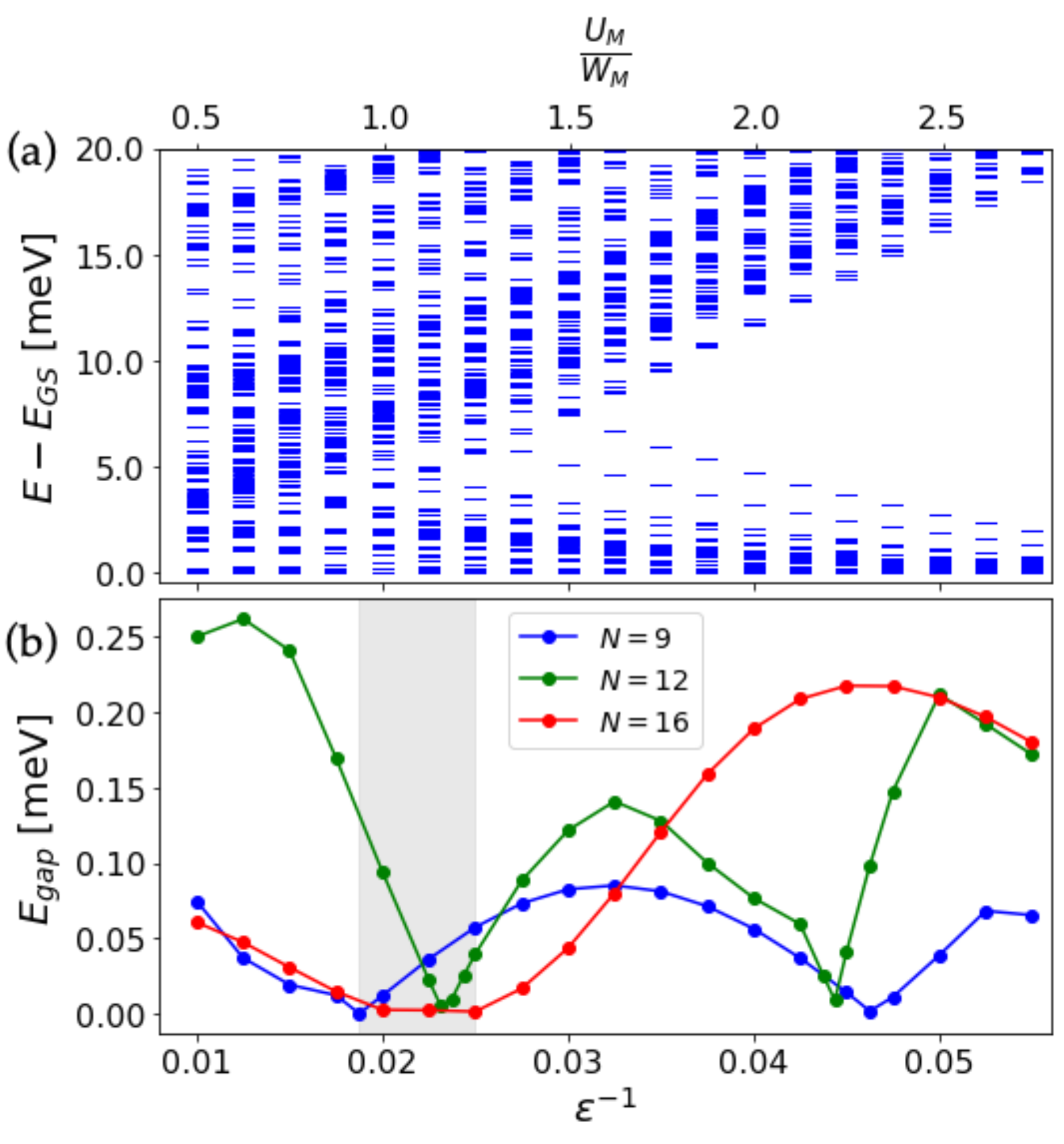}
\caption{(a) The many body spectrum at half-filling for $N=9$ as a function of $\epsilon^{-1}$. The clear separation of two Hubbard bands for strong interactions is visible, with $2^N$ low-energy states. (b) The energy gap between the ground state and the first excited many-body state for systems with $N=9, 12$ and $16$.
The gray-shaded area indicates the region where the metal-insulator transition takes place.  The parameters used
for these calculations were $\theta=2.5^{\circ}$ and $V_m=11$ meV, corresponding to $V_m/W_M$=2.86. The top axis indicates the values of $U_M/W_M$ corresponding to a given $\epsilon^{-1}$.}
\label{fig:EDspectrum}
\end{figure}

Our analysis is based on exact diagonalizations of Eq.\eqref{ManyBodyHamiltonian} with 
periodic boundary conditions applied to different finite system sizes, limiting
the number of momentum points in the discretized first Brillouin zone to $N$. 
We note that for half-filling $N$ is also the number of particles in the spinful system. 
Typical results are illustrated in Fig. \ref{fig:EDspectrum}(a) where we plot the lowest 1700 many-body energies 
relative to the ground state as a function of $\epsilon^{-1}$ for $N=9$. 
We see a set of $2^N$ low-energy states separated by a Hubbard gap to higher states at strong interactions. 
This identifies a parameter range of insulating states where the many-body physics can be described by a spin model.

The picture of localized spins breaks down with decreasing interaction strength 
and a transition to a metallic phase is expected. 
Fig. \ref{fig:EDspectrum}(b) shows the energy gap to the first many-body excited state for system sizes $N=9, 12$ and $N=16$. 
For strong interactions, 
the total spin of the system is size dependent with $S=0$ for $N=12$ and $N=16$, and $S=3/2$ for $N=9$. 
The spectra for $N=9$ and $N=12$ show level crossings around $\epsilon^{-1}\sim 0.045$, signaling a possible
spin liquid intermediate phase with minimum total spin between a Fermi liquid and the strong interaction limit,
as predicted for related models \cite{MITDMRG,Lauchli1,Zaletel}. 
There is a level crossing in the gray-shaded region ($\epsilon^{-1}\sim 0.0175-0.025$)
in all three geometries, that we identify with the MIT.
We have estimated the ratio between onsite Hubbard interaction $U_0$ and the nearest neighbor hopping integral 
$t$ for the shaded region, calculated from our model using Wannierization, obtaining $U_0/t\sim 7.9-9.9$. 
This estimate is consistent with previous studies of the triangular Hubbard model \cite{MITDMRG,Lauchli1,Zaletel}.
For moir\'e materials the precise value of $U_0/t$ is dependent on $\theta$, $V_m$ and $\epsilon$, because of 
longer range hopping and non-standard interaction terms.

To examine the MIT more directly we evaluate the charge gap $\Delta_c$, {\it i.e.}, the difference between the energy to add a particle and the energy to remove a particle from a given ground state, to see if it remains finite 
in the thermodynamic limit. The charge gap shown in Fig.~\ref{fig:Chargegap}(a) 
is defined as $\Delta_c \equiv \lim_{N \to \infty} \Delta_c(N)$, where
\begin{equation}
    \Delta_c(N)=E_0(N+1)+E_0(N-1)-2\, E_0(N).
\end{equation}
The values of $\Delta_c$ for each potential stregth $V_m$ shown in Fig.~\ref{fig:Chargegap} were obtained from extrapolations of $N=4, 9, 16$ results to $N=\infty$ (see Supplemental Material for further details \cite{Supplemental}). The values obtained for the charge gaps in the insulating region of our phase diagram are in the order of tens of 
milielectronvolts, in agreement with results reported in Refs~\onlinecite{HubbardCornell,Columbia}. As noted earlier, these charge gap calculations
show clear metal-insulator phase transitions at positions that can be accurately estimated. The insulating state is favored, as expected, at large potential strengths and at smaller twist angles, which decrease $W_M$ and produce a rapid decrease in band width at a fixed $V_m$. 


\begin{figure}[h!]
\includegraphics[width=\linewidth]{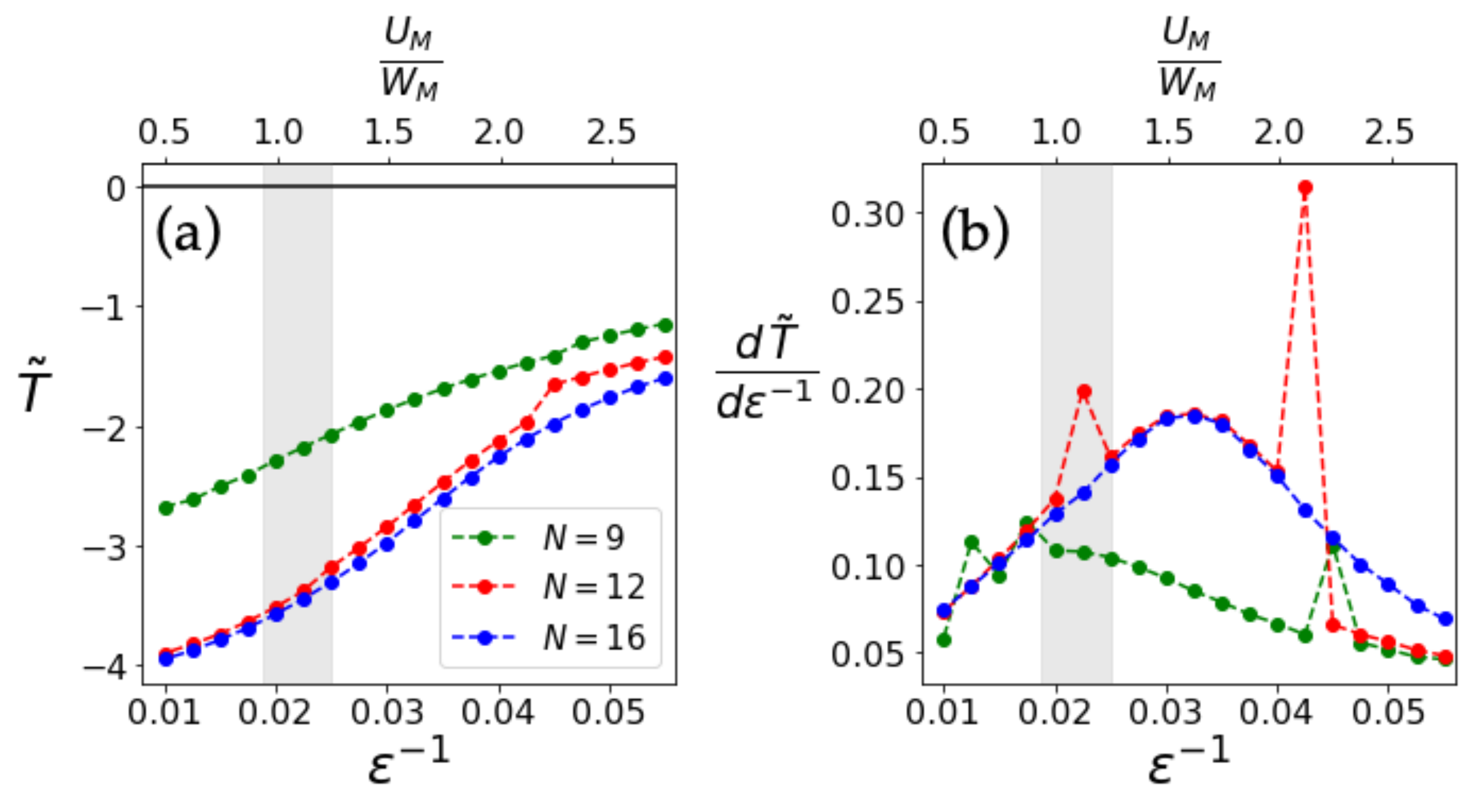}
\caption{(a) Expectation value of single-particle kinetic energy 
relative to the flat-band average $\tilde{T}$ {\it vs}. $\epsilon^{-1}$ 
for several system sizes.  
The gray shading indicates the approximate position of the MIT as estimated by the charge gap calculations.
(b) Numerical first derivative of (a), that reveals peaks for $N=9$ and $N=12$ at both phase transitions. In these calculations $\theta=2.5^{\circ}$ and $V_m=11$ meV, corresponding to $V_m/W_M$=2.86. The top axis indicates the values of $U_M/W_M$ corresponding to a given $\epsilon^{-1}$.}
\label{fig:KineticEnergy}
\end{figure}

Metal-insulator transitions are more interesting when continuous but are usually first-order \cite{KappaET,Tremblay1,Zaletel}.
In magnetically frustrated systems such as the organic compounds \cite{KappaET4,MITDMRG} 
metal-insulator transitions are often only weakly first-order.  Some theoretical work \cite{ContinuousMott1, ContinuousMott2} 
suggests that spin liquids with a 
spinon Fermi surface could undergo continuous metal-insulator phase transitions, with implications for
finite temperature behavior near the critical point. To address the order of the metal-insulator transition
in moir\'e materials, we calculate the expectation value of 
the kinetic energy per particle, relative to the average energy of the band, $\tilde{T}=\langle T\rangle/N-E_{\text{av}}$.
This metallic bonding energy is maximized when interactions are absent and is
expected to be larger in magnitude when the system is more weakly correlated.
If the transition were first-order this quantity would be discontinuous. 
Figure \ref{fig:KineticEnergy}(a) shows the evolution of $ \tilde{T}$ with interaction strength.
At the system sizes we are able to study there is no clear signature of discontinuous behavior,
suggesting that the metal-insulator phase transition in moir\'e materials is either continuous, or only weakly 
first-order. Figure \ref{fig:KineticEnergy}(b) plots the numerical first derivative of $\tilde{T}$ with respect to 
interaction strength. Here we can see peaks for $N=9$ and $N=12$ in the shaded area, while for $N=16$ the derivative seems to be continuous.
This evidence, combined with the apparently continuous vanishing of the charge gap $\Delta_c$ 
{\it vs.} $U_M/W_M$ in Fig.~\ref{fig:Chargegap}(a), clearly shows that the MIT in moir\'e materials 
is not a simple strongly first order phase change. 
\section{\label{sec:level5} Discussion}

The theory of metal-insulator transitions in two or more dimensions continues to be a challenge,
partly because of the absence of a clear order parameter.  In the case of triangular lattice systems,
magnetic frustration in the insulating state adds an additional complication. 
A standard way to approach this problem theoretically
is to study generalized single band Hubbard models 
in particular lattice geometries.
Some layered organic compounds are believed to be described by a triangular lattice Hubbard model with on site interactions and nearest-neighbor 
hopping.  In those systems an intermediate spin-liquid state seems to appear \cite{KappaET,KappaET2,KappaET3,KappaET4} 
in the vicinity of pressure-induced MITs.  Previous numerical studies of the frustrated Hubbard model motivated by these experiments  
do identify the expected insulating ($120^{\circ}$-N\'eel state) and 
the Fermi liquid states in the strong and weak on-site interaction limits \cite{Tremblay1, Tremblay2, MITDMRG,Zaletel}. 
Between those phases an insulating phase without apparent magnetic order 
appears in agreement with experiment, separated from the metal by a first-order transition \cite{Zaletel, Lauchli1, Lauchli2}.
Our calculations suggest that there is also a delicate intermediate state close to the MIT line in triangular lattice moir\'e materials and that the transition occurs   
under experimentally realizable conditions. 
It is clear from our numerical study that the moir\'e material metal-insulator transition is not strongly first order, 
in agreement with known properties of organic compound triangular lattice systems, 
in which magnetic frustration plays an important role. 

The principal difference between moir\'e materials and atomic crystals is that the 
potential that attracts particles to lattice sites is bounded in the former case, 
and unbounded Coulomb ion-core attraction in the latter.  In some cases \cite{LiangFuTransfer,LiangFu2,FengchengTMD1} 
the moir\'e potential can have two minima, and even two-identical minima per moir\'e unit cell \cite{Mattia}.  One signal of this behavior is a relatively small splitting between the two 
topmost moir\'e minibands. In these cases the moir\'e insulator is more like a charge-transfer 
insulator than like a Mott-Hubbard insulator, and the minimal model for the description of its MIT includes at least two-bands.  The boundary between Mott and charge-transfer insulators 
is set by the band structure shape parameter $\psi$, as we show in Fig. \ref{fig:BandStructure}.  

Moir\'e materials are of special interest because of the possibility they present for {\it in situ}, tuning of relevant parameters.
Most important among these is the possibility of using gates to alter the carrier density and to measure the chemical potential 
as a function of carrier density \cite{ChemicalPot1,ChemicalPot2}. 
For metal-insulator phase transitions, the implication 
is that the charge gap at half-filling is directly measurable.  Because the band-width in all heterobilayer moir\'e materials is very sensitive to 
twist angle, this knob can be used to prepare samples that are in the neighborhood of the metal-insulator transition.
{\em In situ} tuning through the metal-insulator phase transition can then be achieved using gates, or pressure, or
by changing gate screening properties.  The prospects for unambiguous experimental determination of the order of the metal-insulator 
phase transition using transport \cite{Drobosavljevic,Tremblay3} and chemical potential measurements,
and of the presence or absence of a spin-liquid state are excellent, and would set the stage for careful studies of 
weakly doped Mott insulators. 

{\it Acknowledgment} -- The authors acknowledge helpful interactions with Naichao Hu, Kin Fai Mak, and Jie Shan. This work was supported by the U.S. Department of Energy, Office of Science, Basic Energy Sciences, under Award $\#$ DE‐SC0019481. PP acknowledges financial support by the Polish National Agency for Academic Exchange (NAWA). We acknowledge the Texas Advanced Computing Center (TACC) at The University of Texas at Austin for providing the high-performance computer resources used for our exact diagonalization calculations.\\

{\it Note added} -- Two experimental  studies  of  the metal-insulator phase transition in  moir\'e superlattices have appeared on the arXiv since our original submission \cite{ExperimentMIT1,ExperimentMIT2}.

\bibliography{refs}


\section*{Comparison between Continuum and Hubbard Coulomb matrix elements}

\begin{figure}[h!]
\includegraphics[width=\linewidth]{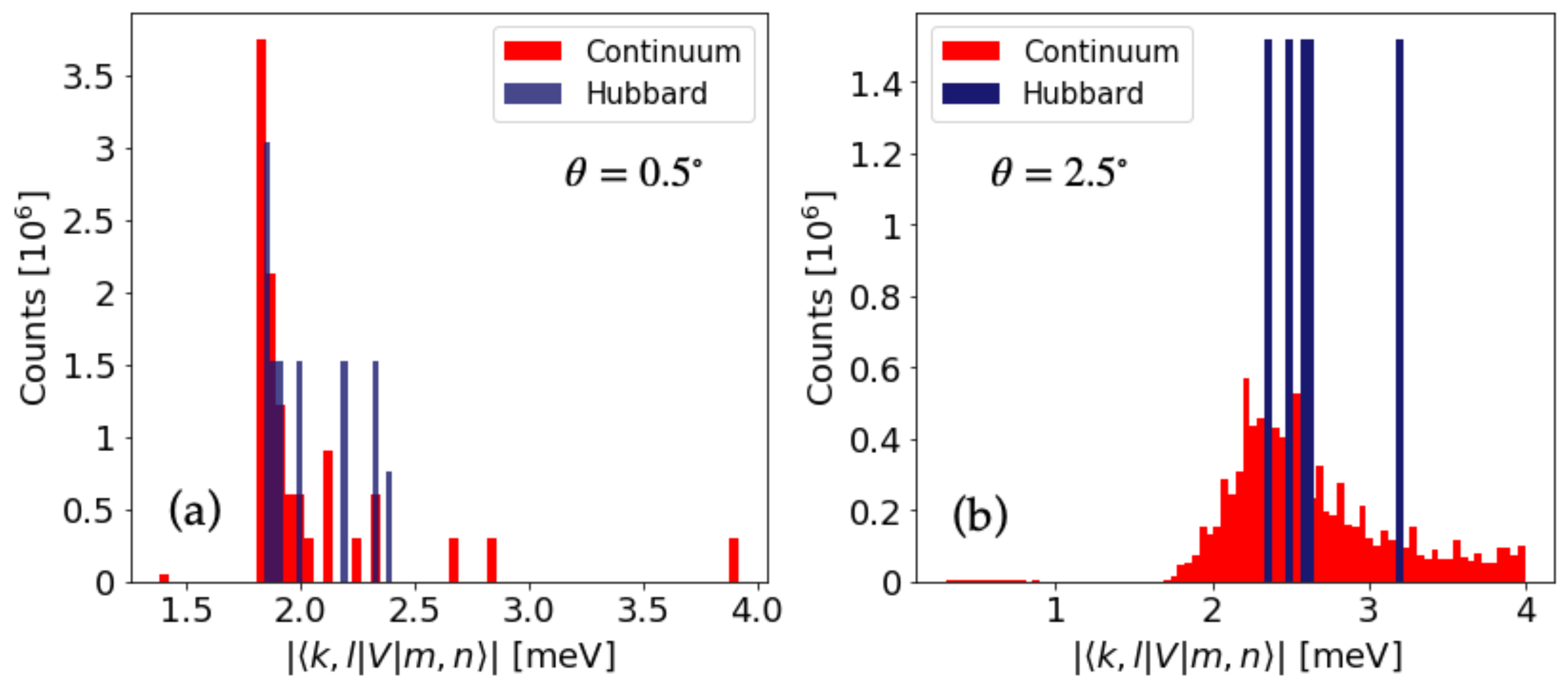}
\caption{Histograms of absolute values of Coulomb matrix elements $|\langle {\bf k},{\bf l}|V|{\bf m},{\bf n}\rangle|$ obtained from WSe$_2$/MoSe$_2$ reciprocal space and triangular Hubbard $U_0+U_1$ models for (a) $\theta=0.5$ and (b) $\theta=1.5$, for $V_m=11$ meV.}
\label{fig:Histograms}
\end{figure}

A comparison between our TMD heterobilayer reciprocal space model and the Hubbard model on a triangular lattice can be made by looking at the distribution of two-body Coulomb matrix elements $\langle {\bf k},{\bf l}|V|{\bf m},{\bf n}\rangle$ as the momentum labels
${\bf k},{\bf l},{\bf m},{\bf n}$ are varied over the Brillouin zone.  To make the comparison between moir\'e materials and 
Hubbard model systems we construct a Wannier function from the topmost moir\'e band wavefunctions and use it to 
calculate $U_0$, and near-neighbor, $U_1$, Hubbard parameters.
Using these parameters, we define a $U_0+U_1$--Hubbard model on a triangular lattice and calculate the distribution 
function of two-particle matrix elements at momentum conserving points in ${\bf k},{\bf l},{\bf m},{\bf n}$-space.
Fig. \ref{fig:Histograms} compares the histograms of the moir\'e material model and the related 
generalized Hubbard model for (a) $\theta=0.5$ and (b) $\theta=2.5$ with $V_m=11$ meV in a Brillouin zone mesh of 225 momentum points.  
For $\theta=0.5$, the histograms have substantial overlap, sharing a peak near $|\langle {\bf k},{\bf l}|V|{\bf m},{\bf n}\rangle|=2.0$ meV. 
If we considered approximating the moir\'e material by an on-site only $U_0$ Hubbard model, 
a single peak would be present near $|\langle {\bf k},{\bf l}|V|{\bf m},{\bf n}\rangle|=1.8$ meV. 
The Wannier-estimation of the nearest-neighbor interaction $U_1$ is an order of magnitude smaller than the 
estimated $U_0$, but for $\theta=0.5$ it spreads the distribution into several peaks and partially explains
the differences compared to the moir\'e material model. 
We conclude that at small twist angles the moir\'e materials Hamiltonian is faithfully represented by a simple Hubbard model, with on-site and perhaps near-neighbor interactions.
For larger angles, {\it i.e.} $\theta=2.5$, the discrepancy between the models is clear, 
suggesting that non-local interaction terms that are normally neglected in lattice models start to play an important role. 
A detailed analysis of the role of long range Coulomb elements and non-local terms in mor\'e materials is a subject of future work.

\section*{Extrapolations to the thermodynamic limit and comparison with Hartree-Fock method}
\begin{figure}[h!]
\includegraphics[width=\linewidth]{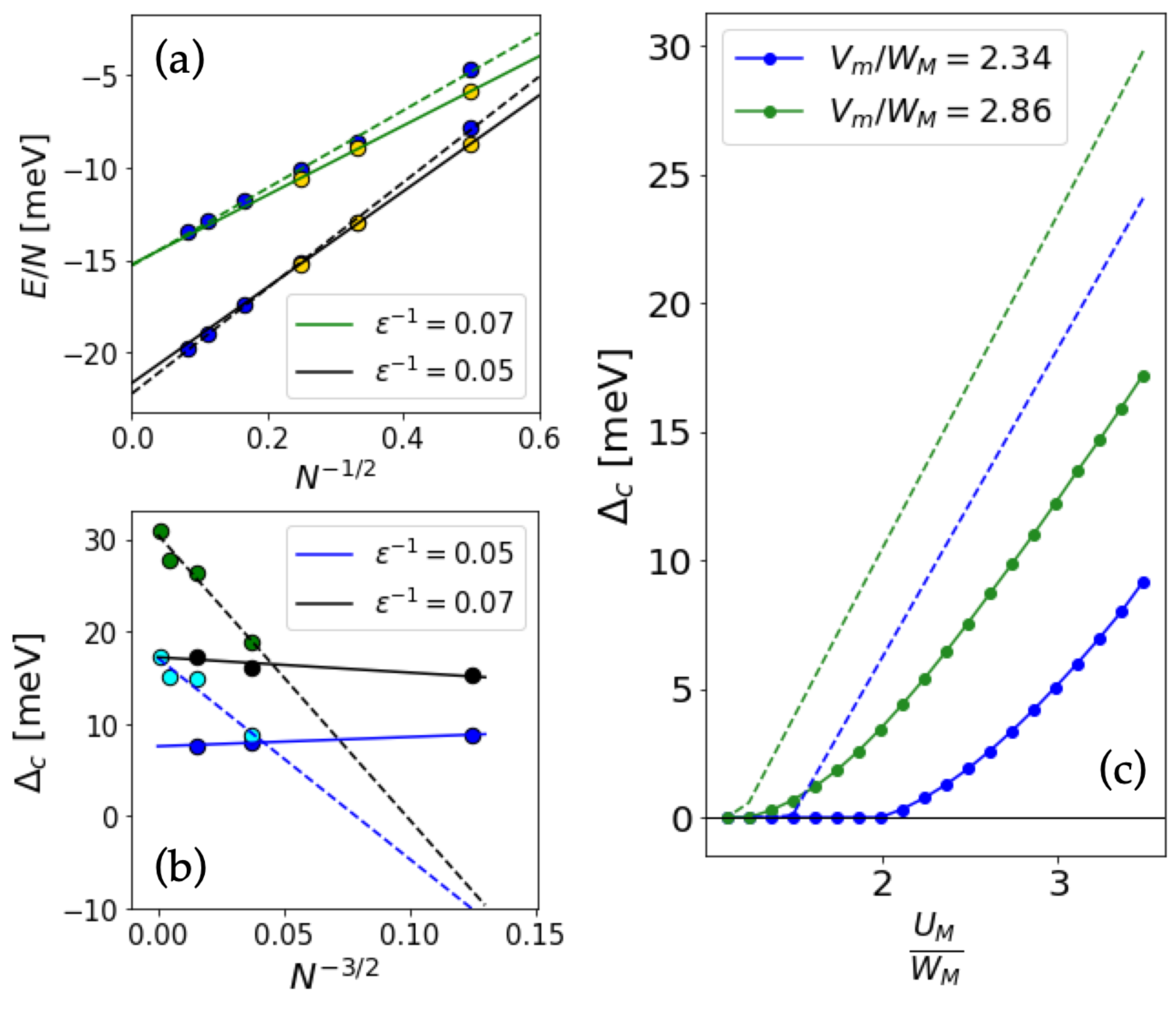}
\caption{System-size extrapolations for exact diagonalization and Hartree-Fock methods. 
(a) Groundstate energies per particle as a function of $N^{-1/2}$ for half-filling at two values of $\epsilon^{-1}$, yellow (blue) points show results from exact diagonalization (Hartree-Fock) for different system sizes. Solid (dashed) lines correspond to extrapolations to the thermodynamic limit from ED (HF) data, for $V_m$=11 meV and $\theta=2.5^{\circ}$. (b) Extrapolations for the charge gaps calculated from ED (solid lines) and Hartree-Fock (dashed lines) for the same parameters as in (a). (c) Extrapolated charge gaps {\it vs} $U_M/W_M$,
for the two values of the normalized moir\'e potential depth $V_m/W_M$ presented in Fig. \ref{fig:Chargegap}(a) (lines with points) and the corresponding charge gaps in a ferromagnetic groundstate calculated from Hartree-Fock (dashed lines).}
\label{fig:Extrapolations}
\end{figure}

We compare extrapolations to larger system sizes for quantities obtained from exact diagonalization
with the corresponding extrapolations for quantities obtained from Hartree-Fock method calculations.
For the Hartree-Fock calculations we assume that translational symmetry is not broken,
{\it i.e.} we assume a ferromagnetic or paramagnetic groundstate,
which allows access to larger system sizes. 
In Fig. \ref{fig:Extrapolations}(a) we show that Hartree-Fock method ground state energy per particle 
results for system sizes $N=4,9, 16, 36, 81$ and $144$ lie in a line when plotted as function of $N^{-1/2}$, as do the exact diagonalization results for system sizes $N=4, 9$ and $16$ discussed in
the main text.  The thermodynamic limit Hartree-Fock energy extrapolated from results for 
$N=4, 9,16$ equals the result extrapolated from calculations at larger system sizes to within less than 
1 meV per unit cell.  We expect that the exact-diagonalization ground state energy results
discussed in the main text have a similar accuracy.
Fig. \ref{fig:Extrapolations}(b) compares charge-gap $\Delta_c \sim N^{-3/2}$ extrapolations 
based on ferromagnetic Hartree-Fock and exact diagonalization calculations.
We see that the results follow the expected power laws accurately. In Fig. \ref{fig:Extrapolations}(c) we show how the charge gaps obtained from Hartree-Fock compare to the exact diagonalization charge gaps as a function of interaction strength. As pointed out in the MS, Hartree-Fock mean-field method tends to overestimate the stability of the insulating phase, which is manifested in larger values for $\Delta_c$ and also 
in metal-insulator transition points at smaller values of $U_M/W_M$. 
This discrepancy between charge gaps justifies the necessity of using a non-perturbative method like exact diagonalization in order to study physics near the metal-insulator phase transition. 

\end{document}